\documentclass[aps,prl,twocolumn,showpacs,superscriptaddress,groupedaddress]{revtex4}  
\usepackage{graphicx}  
\usepackage{dcolumn}   
\usepackage{bm}        
\usepackage{amssymb}   
\usepackage{multirow}
\usepackage{color}

\begin{document}



\title{Determination of the Axial-Vector Weak Coupling Constant with Ultracold Neutrons
}

%
\author{J.~Liu}
\affiliation{Kellogg Radiation Laboratory, California Institute of Technology, Pasadena, California 91125, USA}
\affiliation{Department of Physics, Shanghai Jiao Tong University, Shanghai, 200240, China}
\author{M.~P.~Mendenhall}
\affiliation{Kellogg Radiation Laboratory, California Institute of Technology, Pasadena, California 91125, USA}
\author{A.~T.~Holley}
\affiliation{Department of Physics, North Carolina State University, Raleigh, North Carolina 27695, USA}
\author{H.~O.~Back}
\affiliation{Department of Physics, North Carolina State University, Raleigh, North Carolina 27695, USA}
\affiliation{Triangle Universities Nuclear Laboratory, Durham, North Carolina 27708, USA}
\author{T.~J.~Bowles}
\affiliation{Los Alamos National Laboratory, Los Alamos, New Mexico 87545, USA}
\author{L.~J.~Broussard}
\affiliation{Triangle Universities Nuclear Laboratory, Durham, North Carolina 27708, USA}
\affiliation{Department of Physics, Duke University, Durham, North Carolina 27708, USA}
\author{R.~Carr}
\affiliation{Kellogg Radiation Laboratory, California Institute of Technology, Pasadena, California 91125, USA}
\author{S.~Clayton}
\affiliation{Los Alamos National Laboratory, Los Alamos, New Mexico 87545, USA}
\author{S.~Currie}
\affiliation{Los Alamos National Laboratory, Los Alamos, New Mexico 87545, USA}
\author{B.~W.~Filippone}
\affiliation{Kellogg Radiation Laboratory, California Institute of Technology, Pasadena, California 91125, USA}

\author{A.~Garc\'{\i}a}
\affiliation{Department of Physics, University of Washington, Seattle, Washington 98195, USA}
\author{P.~Geltenbort}
\affiliation{Institut Laue-Langevin, 38042 Grenoble Cedex 9, France}
\author{K.~P.~Hickerson}
\affiliation{Kellogg Radiation Laboratory, California Institute of Technology, Pasadena, California 91125, USA}
\author{J.~Hoagland}
\affiliation{Department of Physics, North Carolina State University, Raleigh, North Carolina 27695, USA}
\author{G.~E.~Hogan}
\affiliation{Los Alamos National Laboratory, Los Alamos, New Mexico 87545, USA} 
\author{B.~Hona}
\affiliation{Department of Physics and Astronomy, University of Kentucky, Lexington, Kentucky 40506, USA}
\author{T.~M.~Ito}
\affiliation{Los Alamos National Laboratory, Los Alamos, New Mexico 87545, USA}
\author{C.-Y.~Liu}
\affiliation{Department of Physics, Indiana University, Bloomington, Indiana 47408, USA}
\author{M.~Makela}
\affiliation{Los Alamos National Laboratory, Los Alamos, New Mexico 87545, USA}
\author{R.~R.~Mammei}
\affiliation{Department of Physics, Virginia Tech, Blacksburg, Virginia 24061, USA}
\author{J.~W.~Martin}
\affiliation{Department of Physics, University of Winnipeg, Winnipeg, MB R3B 2E9, Canada}
\author{D.~Melconian}
\affiliation{Cyclotron Institute, Texas A$\&$M University, College Station, Texas 77843, USA}
\author{C.~L.~Morris}
\affiliation{Los Alamos National Laboratory, Los Alamos, New Mexico 87545, USA}
\author{R.~W.~Pattie Jr.}
\affiliation{Department of Physics, North Carolina State University, Raleigh, North Carolina 27695, USA}
\affiliation{Triangle Universities Nuclear Laboratory, Durham, North Carolina 27708, USA}
\author{A.~P{\'e}rez Galv{\'a}n}
\affiliation{Kellogg Radiation Laboratory, California Institute of Technology, Pasadena, California 91125, USA}
\author{M.~L.~Pitt}
\affiliation{Department of Physics, Virginia Tech, Blacksburg, Virginia 24061, USA}
\author{B.~Plaster}
\affiliation{Department of Physics and Astronomy, University of Kentucky, Lexington, Kentucky 40506, USA}
\author{J.~C.~Ramsey}
\affiliation{Los Alamos National Laboratory, Los Alamos, New Mexico 87545, USA}
\author{R.~Rios}
\affiliation{Los Alamos National Laboratory, Los Alamos, New Mexico 87545, USA}	
\affiliation{Department of Physics, Idaho State University, Pocatello, Idaho 83209, USA}
\author{R.~Russell}
\affiliation{Kellogg Radiation Laboratory, California Institute of Technology, Pasadena, California 91125, USA}
\author{A.~Saunders}
\affiliation{Los Alamos National Laboratory, Los Alamos, New Mexico 87545, USA}
\author{S.~J.~Seestrom}
\affiliation{Los Alamos National Laboratory, Los Alamos, New Mexico 87545, USA}
\author{W.~E.~Sondheim}
\affiliation{Los Alamos National Laboratory, Los Alamos, New Mexico 87545, USA}
\author{E.~Tatar}
\affiliation{Department of Physics, Idaho State University, Pocatello, Idaho 83209, USA}	
\author{R.~B.~Vogelaar}
\affiliation{Department of Physics, Virginia Tech, Blacksburg, Virginia 24061, USA}
\author{B.~VornDick}
\affiliation{Department of Physics, North Carolina State University, Raleigh, North Carolina 27695, USA}
\author{C.~Wrede}
\affiliation{Department of Physics, University of Washington, Seattle, Washington 98195, USA}
\author{H.~Yan}
\affiliation{Department of Physics and Astronomy, University of Kentucky, Lexington, Kentucky 40506, USA}
\author{A.~R.~Young}
\affiliation{Department of Physics, North Carolina State University, Raleigh, North Carolina 27695, USA}
\affiliation{Triangle Universities Nuclear Laboratory, Durham, North Carolina 27708, USA}

%
\collaboration{The UCNA Collaboration}
\vskip 0.25cm

\date{\today}

\begin{abstract}
A precise measurement of the neutron decay $\beta$-asymmetry $A_0$ has
been carried out using polarized ultracold neutrons (UCN) from the
pulsed spallation UCN source at the Los Alamos Neutron Science
Center (LANSCE). Combining data obtained in 2008 and 2009, we report
$A_0 = -0.11966 \pm 0.00089 _{-0.00140}^{+0.00123}$, from which we
determine the ratio of the axial-vector to vector weak coupling of the
nucleon $g_A/g_V = -1.27590 _{-0.00445}^{+0.00409}$.
\end{abstract}

\maketitle




The axial-vector weak coupling constant, $g_A$, plays an
important role in our understanding of the nucleon spin and flavor 
structure~\cite{bass05, close88}. 
It is a central target for high precision lattice
QCD calculations~\cite{yamazaki08,choi10} and an essential parameter
in effective field theories~\cite{gockeler05}.
$g_A$ is also important in a variety of astrophysical
processes, including solar fusion reaction rates~\cite{adelberger10}.



The angular distribution of emitted electrons from polarized neutron
decay can be expressed as $W(E)\propto 1 + \frac{v}{c} \langle P \rangle
A(E) \cos\theta$, where $A(E)$ specifies the $\beta$-asymmetry versus
electron energy $E$, 
$v$ is the electron velocity, $c$ is the speed of light,
$\langle P \rangle$ is the mean polarization, and $\theta$ is the angle between
the neutron spin and the electron emission direction
\cite{jackson57}. The leading order value of $A(E)$, $A_0$, is given
by
\begin{equation}
  A_0 = \frac{-2(\lambda ^2 - |\lambda|)}{1+3\lambda ^2},
\label{eq:A0_lambda}
\end{equation}
where $\lambda = g_A/g_V$ and $g_V$ is the vector weak coupling constant with $g_V = 1$ under the conserved vector current (CVC) hypothesis of the Standard Model \cite{towner10}. 
Higher order terms in $A(E)$ are at the 1\% level, and can be 
calculated precisely under the Standard Model~\cite{wilkinson82,gardner01}.
$g_A$ can also be indirectly determined  
by combining the Fermi coupling constant $G_F$, 
measured to 5 ppm using muon decay \cite{chitwood07}, the CKM matrix element
$|V_{ud}|$, measured to 225 ppm using $0^+ \rightarrow 0^+$
superallowed decays \cite{towner10}, and the neutron lifetime, measured 
to 0.9\% \cite{PDG} \cite{czarnecki04}. Thus, a
measurement of the $\beta$-asymmetry 
permits direct determination of $g_A$, as well as a robust test of 
the consistency of measured neutron
$\beta$-decay observables under the Standard Model.

In order to obtain $A(E)$, one must determine the polarization of the
neutron beam and control all sources of systematic uncertainty due to
backgrounds, including those produced by the neutrons themselves, the
detector response, and electron-event reconstruction. 
All previous precise measurements of the $\beta$-asymmetry \cite{abele02, liaud,
yerozlim, bopp} have been performed 
with cold-neutron beams and have shown a range 
of results much wider than the reported uncertainties \cite{PDG}.
Our measurement, UCNA,
utilizes ultracold neutrons 
(neutrons with kinetic energy less than 200 neV)
and
controls key systematic uncertainties: 
neutron polarization and neutron-generated backgrounds. 
In 2007, we carried out 
a proof-of-principle $\beta$-asymmetry measurement~\cite{ucna_prl09}.
At present, the UCNA experiment is characterized by
neutron polarizations greater than 99.48\%
and neutron-generated
backgrounds that produce corrections to the asymmetry below the 0.02\%
level.

Some of the experimental details of UCNA are explained in~\cite{ucna_prl09}.
We used the UCN source at the LANSCE accelerator at Los Alamos National
Laboratory~\cite{source_pl04}. The UCN were polarized by 
a 7~T primary polarizer coupled to an
adiabatic fast passage (AFP) spin flipper to control the spin
state~\cite{ucna_prl09,afp_nim}.  Polarized UCN entered the superconducting
spectrometer (SCS)~\cite{scs_nim} and were confined in a 3 m long,
12.4 cm diameter electropolished Cu tube (decay trap) with variable
thickness mylar endcaps.  The inside surface of each endcap was coated
with 200 nm of Be.  A 1~T magnetic field was oriented parallel to the
decay trap, along which decay electrons spiraled toward one of two
identical electron detector packages, each covering a $2\pi$ sr angular
hemisphere.
Each detector package
consisted of a low-pressure multiwire proportional chamber (MWPC) 
\cite{mwpc_nim} backed
by a plastic scintillator, with scintillation light measured by four 
photomultiplier tubes (PMT). 
Each MWPC had thin front and back mylar windows which separated
low-Z chamber gas (neopentane) from the spectrometer vacuum.  To study key
systematics due to electron energy loss and backscattering in the
windows, we operated the experiment in four different geometries
with different decay trap endcaps and MWPC window 
thicknesses, A: 0.7 and 25, B: 13.2 and 25, C:  0.7 and 6, and 
D~\footnote{The entrance guide for Geometry D 
was diamond-like-carbon-coated
Cu, as compared to bare Cu for the other geometries, resulting in a
different UCN velocity spectrum in the decay trap.}: 0.7 and 6 $\mu$m, respectively.


Cosmic-ray muon backgrounds were identified by a combination of 
plastic scintillator veto paddles and sealed
drift tube assemblies~\cite{drift_tube_nim} surrounding the electron detectors.
$\gamma$-ray backgrounds were vetoed by a coincidence 
between the MWPC and the main $\beta$-scintillator.

\begin{figure}
\includegraphics[width=0.4\textwidth]{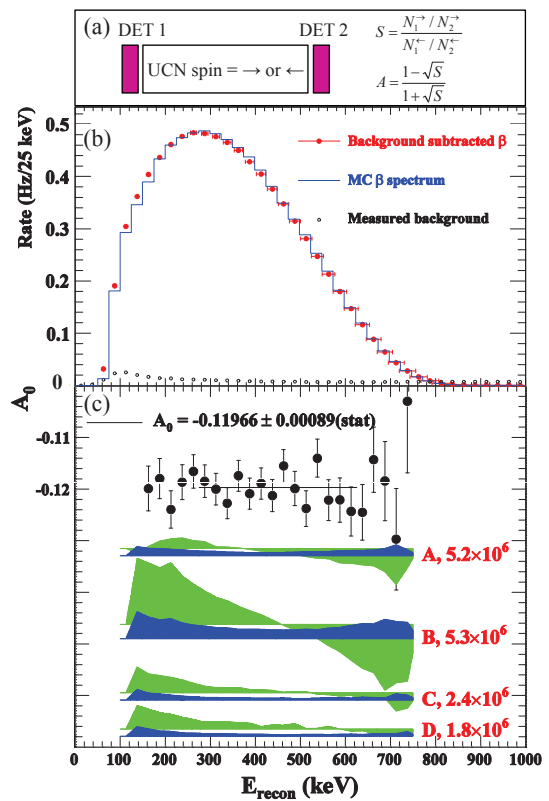}
\caption{(Color online)
  Panel (a): schematic of the experiment and definition of the asymmetry. 
  Panel (b): background subtracted electron $E_{\mathrm{recon}}$ spectrum (solid
  circles
  with the uncertainty of $E_{\mathrm{recon}}$ reflected by 
  the horizontal error bars), 
  combining both sides of the detectors and two spin states, overlaid 
  with the Monte Carlo spectrum (histogram). The open 
  circles represent the measured 
  ambient 
  background spectrum.  Panel (c): $A_0$ vs.\
  $E_{\mathrm{recon}}$, combining all four geometries.
  The horizontal line represents the extracted $A_0$ within 
  [275, 625] keV. 
  Drawn on the same scale below the graph are four sets of 
  bands representing the sum of energy-dependent 
  backscattering and angle effect corrections (light color), 
  positive sign indicating a larger $|A_0|$, and their uncertainties (dark) 
  for the four geometries.
  The positive (negative) correction at low (high) energy is
  a consequence of the backscattering (angle effect) dominating in this
  energy region. 
  The number of $\beta$ events for each geometry is also 
  indicated in the figure.
}
\label{fig:energy_spec}
\end{figure}

A gate valve separated the UCN
source from the experimental apparatus. A typical run unit consisted of
a background run (gate valve closed), a $\beta$-decay run (gate valve
open), and a UCN depolarization run to measure the 
equilibrium UCN polarization for the accompanying $\beta$-decay run.
The UCN spins were flipped back and forth (while the magnetic 
field in the SCS was held fixed) between run units
(Fig.~\ref{fig:energy_spec}, Panel (a)), 
which partially canceled 
systematic rate variation over the period of a spin cycle 
($\sim$1.5 hour).
During a depolarization run,  the guide serving as input to the 
7~T polarizing field was first connected to a
UCN detector~\cite{morris09} so that UCN exiting the experiment 
could be counted, while the gate valve was closed and the proton 
pulses~\cite{source_pl04}
were discontinued.
This \textit{cleaning} phase, 
which lasted 25~s, produced a signal in the UCN detector proportional 
to the number of correctly polarized UCN present in the experimental 
geometry at the end of the $\beta$-decay measurement interval. Following 
the cleaning phase, the state of the spin flipper was changed, 
allowing only
incorrectly polarized UCN remaining downstream of the spin flipper 
to pass through the 7~T polarizing field and be counted. Counting 
during this \textit{unloading} phase was performed for $\sim$200~s
in order to measure background as well as incorrectly polarized UCNs.
Since the measured depolarization was consistent with zero at the 
$1\sigma$ level, we folded together statistical and systematic errors 
to produce a global polarization lower limit of 99.48\% at the 
68\% CL~\cite{afp_nim}, covering all four geometries and both 
polarization states.

The experimental triggers were formed by requiring
at least 2-of-4 PMT signals over threshold in either of the 
scintillator detectors.  
Electron positions were determined with the MWPC to an accuracy of
better than 2 mm
based on the distribution of charge on two perpendicular 
cathode grids in the MWPC~\cite{mwpc_nim}. 
A fiducial cut of $r<45$~mm 
was placed on the trigger side to reduce background 
and to eliminate electrons that could strike the decay trap
walls.  

Reconstructed event energies $E_{\mathrm{recon}}$ were measured using
the signals from the scintillator PMTs as calibrated with conversion
electron sources ($^{109}$Cd, $^{139}$Ce, $^{113}$Sn, $^{85}$Sr, and
$^{207}$Bi).  The position-dependence of the response of each PMT was
mapped out by comparing the neutron beta decay spectrum endpoint
observed at different positions.  
The energy reconstruction uncertainty was determined to be the larger of $\pm
5$~keV or $\pm 2.5$\%, which covered the uncertainty in the position 
response of the scintillator, as well as possible variation of the 
energy response allowed by the calibration data.

The PMT gains were monitored based on frequent calibrations with a
removable ${}^{113}$Sn calibration source, which also measured the
energy resolution of the system ($\sim400$ photoelectrons per MeV), and
by observing shifts in the minimum-ionizing peak of cosmic-ray muons
during $\beta$-decay and background runs.

The majority of the $\beta$-decay events were single detector triggers. 
However, due to electron backscattering, a small fraction of the events,
varied between 1.7\% and 3.4\% for the four geometries,
triggered both scintillators and another small fraction ($\sim$2\%) 
were detected by both
MWPCs, but triggered only one of the scintillators. 
In the first case, the initial direction of the electron 
could be determined by the relative timing of the triggers, while in the
second case 
a fixed cut (4.1 keV) or a likelihood function based on the energy loss in the 
trigger side MWPC yielded an identification 
efficiency of $\sim$80\% based on Monte Carlo calculations (discussed later).
In addition to the ambient backgrounds (measured with the 
UCN gate valve closed and suppressed by the pulsed nature 
of the UCN source~\cite{source_pl04}), 
which were subtracted run-by-run, neutron captures
in the vicinity of the detectors could create prompt $\gamma$'s with
energies up to $\sim8$ MeV, generating an irreducible background in the
experiment. This background was significantly suppressed, compared to 
cold neutron beam experiments, by the relatively low density 
and low capture and upscatter probability of neutrons in and 
around the spectrometer.
Combining direct measurements with Monte Carlo calculations, we
obtained an upper limit of 0.02\% on the correction to the asymmetry.

For each run, events were sorted into 25~keV $E_{\mathrm{recon}}$ 
bins from 0 to 1200 keV and assigned an initial 
direction.
The rates in the two detectors were then computed based on the
experiment live time.
We applied separate spin-dependent blinding factors to the two
detector rates, effectively adding an unknown scaling factor
to the measured asymmetry that was constrained to be within $1.00 \pm
0.05$.  After determination of all cuts, corrections and
uncertainties, this factor was removed.  For each
$\beta$-decay/background run pair, the background rate was subtracted
from the $\beta$-decay-run rate bin by bin.  The reconstructed
energy spectrum (background subtracted, averaged over the two spin
states) is shown in Panel (b) of Fig.~\ref{fig:energy_spec}, 
overlaid with the measured
background. 
The $S/B$ is about 40 in our analysis energy  
window between 275 and 625 keV (discussed later).
Also overlaid is
the Monte Carlo-predicted reconstructed energy spectrum, with all
detector effects (efficiencies, resolutions, etc.) taken into account.
The systematic 
effect due to the small discrepancy between the two spectra is 
well covered by the energy reconstruction uncertainty in Table~\ref{tab:sys}.

For a given geometry, a ``super-ratio'' of count rates among the two detectors 
and UCN spin states was calculated (as defined in \cite{ucna_prl09}), from which 
the raw measured asymmetry was determined (see also Fig.~\ref{fig:energy_spec},
Panel (a)). 
To extract $A_0$, we first multiplied the raw measured asymmetry by
$1/\langle v/c \rangle$ in each energy bin to remove the strongest
energy dependence. As in \cite{ucna_prl09}, two scattering-related
effects dominated subsequent systematic corrections: the residual
backscattering correction and the angle effect.  In addition to a
small residual correction due to incorrect identification of the
initial electron direction for the measured electron backscatters
(where both detectors observed the electron), there were corrections for
backscattering from the decay trap windows and the front
windows of the MWPC that could not be identified experimentally.  Angle
effects arose from the fact that the energy loss of an electron in the
thin windows was strongly angle-dependent.  Low-energy, large pitch
angle electrons were more likely to fall below the scintillator
threshold, leading to a suppression of the acceptance at large angles
($\langle \cos\theta \rangle$ deviating from $1/2$).
Both of these effects were evaluated with two independent simulation
programs: Penelope~\cite{penelope} and GEANT4~\cite{geant4}.  
The resulting corrections for all four geometries are shown in Table~\ref{tab:sys}. 
Based on the observed difference between the calculations and the data, we
assigned an uncertainty of 30\% to the backscattering
correction and 25\% to the angle effect correction.


Recoil-order corrections to $A(E)$ 
(see also \cite{ucna_prl09}) 
were calculated within the context
of the Standard Model according to the formalism of
\cite{wilkinson82,gardner01}, 
leading to a correction
of $-1.79\pm0.03$\% to $A_0$. 
The value for the radiative correction to $A_0$ was taken from the
calculations of \cite{gluck92}, 
yielding a small 
theoretical correction of $0.10\pm0.05$\%.

Applying all corrections mentioned above, the extracted $A_0$ is
plotted against $E_{\mathrm{recon}}$ (all geometries combined) in
Panel (c) of Fig~\ref{fig:energy_spec}. Energy-dependent corrections
(backscattering and angle effects) and their uncertainty are indicated
as bands in the figures. The final $A_0$ is obtained from a constant 
fit over a range of energy~\cite{stump01}. 
The energy window, 275 to 625 keV, was chosen 
to optimize combined statistical and systematic uncertainties
before unblinding the asymmetries. 
The value of $A_0$ was insensitive to the choice of energy window, with 
the variation less than 15\% of the statistical uncertainty for windows 
between 150 and 750 keV.

The experimental uncertainties and systematic corrections to $A_0$ 
are summarized in Table~\ref{tab:sys}. 
Geometry-dependent systematic uncertainties (backscattering, angle
effect, and MWPC inefficiency) are treated as completely correlated
among the different geometries.  Combining the four geometries,
we find $A_0 = -0.11966 \pm 0.00089 _{-0.00140}^{+0.00123}$ 
(with a $\chi^2/\nu$ of $2.4/3$), where the first uncertainty is
statistical and the second systematic~\cite{stump01}.
Based 
on  Eq.\
(\ref{eq:A0_lambda}), we then
determine $g_A/g_V = -1.27590 _{-0.00445}^{+0.00409} = g_A$, where the 
second equality assumes CVC~\cite{towner10}.



\begin{table}[t]
\caption{\label{tab:sys} Summary of 
experimental 
corrections and uncertainties 
in \% (all fractional to $A_0$). Upper: geometry-independent effects.
Lower: geometry-dependent effects (first value=correction, second value=uncertainty in each column), 
with $\sigma_{\mathrm{stat}}$,
$\Delta_{\mathrm{back}}$, $\Delta_{\mathrm{ang}}$ and
$\epsilon_{\mathrm{MWPC}}$ referring to statistical uncertainty,
backscattering correction, angle effect correction, and the uncertainty 
associated with MWPC inefficiency, respectively.}
\addtolength{\tabcolsep}{2.5 mm}
\begin{tabular}{lcc}
\hline\hline
Geometry-independent effect & corr. (\%) & unc. (\%)\\\hline
Polarization & 0 & $\displaystyle_{-0}^{+0.52}$\\
Field non-uniformity & 0 & $\displaystyle_{-0}^{+0.20}$\\
Rate dependent gain shift & 0 & 0.08\\
Gain fluctuation & 0 & 0.20\\
Deadtime & 0 & 0.01\\
Energy reconstruction & 0 & 0.47\\
UCN-induced background & 0 &0.02\\
Muon veto efficiency & 0 & 0.30\\
Live time uncertainty & 0 &0.24\\
Fiducial cut & 0 & 0.24\\
\hline
\end{tabular}
\\
\addtolength{\tabcolsep}{-2.7 mm}
{\small
\begin{tabular}{c|cc|cc|cc|cc}
\multicolumn{9}{c}{Geometry-dependent effect}\\\hline
& \multicolumn{2}{c|}{A (\%)} &\multicolumn{2}{c|}{B (\%)} 
&\multicolumn{2}{c|}{C (\%)} &\multicolumn{2}{c}{D (\%)}\\\hline


$\sigma_{\mathrm{stat}}$ &  n/a & 1.23 & n/a & 1.22 & n/a & 2.00 & n/a & 2.10 \\
$\Delta_{\mathrm{back}}$ & 1.34& $\pm0.40$ & 4.32& $\pm1.30$ & 1.07& $\pm0.32$ & 1.08& $\pm0.32$  \\
$\Delta_{\mathrm{ang}}$ & $-1.81$& $\pm0.45$ & $-3.22$& $\pm0.81$ & $-0.60$& $\pm0.15$ & $-0.36$& $\pm0.09$\\
$\epsilon_{\mathrm{MWPC}}$ & 0 & 0.02 & 0 & 0.01 & 0 & 0.16 & 0 & 0.5\\\hline\hline

\end{tabular}
}
\end{table}


\begin{figure}[!htbp]
  \centering
  \includegraphics[width=0.48\textwidth]{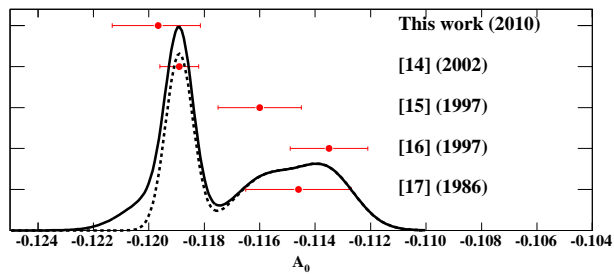}
  \caption{(Color online) Ideogram comparing the previous measurements that are 
    included in the 2009 Particle Data Group best value for $A_0$ 
    along with the new result reported here. The dashed line is the ideogram
    with the previous four data points, and the solid line shows the 
    result of including the present data. 
}
  \label{fig:ideogram}
\end{figure}


Our result for $A_0$ is compared with the world data
\cite{abele02,liaud,yerozlim,bopp} in Fig.\ \ref{fig:ideogram}.
Our result is in good agreement with the most recent and 
precise result for $A_0$~\cite{abele02}.
We note that the direct extraction of $g_A/g_V$ from the $\beta$-asymmetry 
is, unlike extraction from the neutron lifetime~\cite{PDG}\footnote{The 
unresolved 
discrepancy between the most recent lifetime measurement and the previous
is discussed in the summary of neutron properties in \cite{PDG}.}, 
independent of the CKM matrix element $|V_{ud}|$.
This strongly motivates UCN-based measurements of the neutron 
$\beta$-asymmetry, where the key neutron-related systematic uncertainties 
can be reduced below the 0.1\% level. 
Recently demonstrated improvements to the UCN source and refinement 
of the energy response and gain monitoring will permit the collection
of a much larger data set and the reduction of all major systematic 
uncertainties.


%
This work was supported in part by the Department of Energy Office of Nuclear Physics, National
Science Foundation (NSF-0555674, NSF-0855538, NSF-0653222), 
and the Los Alamos National Laboratory LDRD program. We gratefully acknowledge the support of LANSCE and AOT divisions of Los Alamos National Lab.

\end{document}